\documentstyle[emulateapj,psfig,epsfig]{article}
 \begin{document}
 \received{}
 \accepted{}
 \revised{}
 \lefthead{Pritzl et al.}
 \righthead{Oosterhoff classification of NGC~6388 and NGC~6441}
 \slugcomment{ApJ (Letters), in press}
 \title{RR Lyrae Stars in NGC~6388 and NGC~6441:\\A New Oosterhoff Group?}
 \author{Barton Pritzl\altaffilmark{1} and Horace A. Smith}
 \affil{Dept. of Physics and Astronomy, Michigan State University,
 East Lansing, MI 48824 \\$\,$}
 \authoremail{pritzl@pa.msu.edu, smith@pa.msu.edu}

 \author{M\'arcio Catelan\altaffilmark{2}}
 \affil{University of Virginia, Department of Astronomy, P.O.~Box~3818,
 Charlottesville, VA 22903-0818}
 \authoremail{catelan@virginia.edu}

 \and

 \author{Allen V. Sweigart}
 \affil{NASA Goddard Space Flight Center, Laboratory for Astronomy and
 Solar Physics, Code 681,\\Greenbelt, MD 20771}
 \authoremail{sweigart@bach.gsfc.nasa.gov}
 
 \altaffiltext{1}{Visiting Astronomer, Cerro Tololo Inter-American
 Observatory, National Optical Astronomy Observatories, which is
 operated by AURA, Inc., under cooperative agreement with the
 National Science Foundation}
 \altaffiltext{2}{Hubble Fellow.}
 
 \begin{abstract}
 NGC~6388 and NGC~6441 are anomalies among Galactic globular clusters
  in that they cannot be readily placed into either Oosterhoff group I
 or Oosterhoff group II despite their significant numbers of RR Lyrae
 variables.  The mean pulsation periods, $\langle P_{\rm ab} \rangle$,
 of their RRab variables,
 at 0.71~d and 0.76~d, respectively, are even larger than for
 Oosterhoff II clusters.  Moreover, Oosterhoff II
 clusters are very metal-poor, whereas NGC~6388 and NGC~6441 are the most
 metal-rich
 globular clusters known to contain RR Lyrae stars. The location of the
 NGC~6388 and NGC~6441 RRab variables in the period-amplitude diagram
 implies that the RR Lyrae stars in those two clusters are brighter
 than expected for their metallicities.  Our results therefore indicate
 that a universal
 relationship may not exist between the luminosity and the metallicity of RR
 Lyrae
 variables.
 
 \end{abstract}
 

 \section{Introduction}
 
 Oosterhoff (1939)\markcite{Oo39} called attention to a dichotomy in the
 properties of RR Lyrae stars (RRLs) belonging to five RR Lyrae-rich
 globular clusters.  The five clusters could be divided into what are
 now known as Oosterhoff groups I (Oo~I) and II (Oo~II) on the basis
 of the mean periods and relative proportions
 of their RRab and RRc stars (Table~1). Subsequent
 investigations confirmed that all Galactic globular clusters which
 contain significant numbers of RRLs could be assigned to either
 Oo~I or Oo~II.  It also became clear that globular clusters of
 Oo~I were more metal rich than those of Oo~II (Smith 1995\markcite{S95}
 and references therein).  The cause
 of the Oosterhoff dichotomy, and its implications for the brightnesses
 of RRLs and the ages of globular clusters, remains a subject of debate
 (van Albada \& Baker 1973\markcite{V73}; Sandage, Katem, \& Sandage
 1981\markcite{S81}; Castellani 1983\markcite{C83}; Renzini 1983\markcite{R83};
 Lee, Demarque, \& Zinn 1990;
 Sandage 1993a, 1993b\markcite{Sa93a}; Clement \& Shelton 1999\markcite{C99}).
 
 Although RRLs more metal rich than ${\rm [Fe/H]} = -0.8$ are known to
 exist in the field population of the Galaxy (Preston 1959\markcite{P59};
 Layden 1994\markcite{L94}), very few RRLs have been discovered
 within the most metal rich globular clusters.  Metal-rich clusters
 have stubby horizontal branches (HBs) which lie entirely or almost entirely
 to the red side of the instability strip.  The globular clusters
 NGC~6388 and NGC~6441 are prominent exceptions to this rule. These are
 relatively metal-rich globular clusters with ${\rm [Fe/H]} = -0.60$ and
 $-0.53$,
 respectively (Armandroff \& Zinn 1988\markcite{A88}).  Their
 HBs nonetheless have strong blue as well as red components (Rich
 et al. 1997)\markcite{R97}. Studies of NGC~6388 by Hazen \& Hesser
 (1986)\markcite{H86} and Silbermann et al. (1994)\markcite{S94} and of
 NGC~6441 by Layden et al. (1999)\markcite{L99} indicated that both clusters
 contain significant numbers of RRLs. These studies, especially that of
 Layden et al., also showed that RRLs within NGC~6388 and NGC~6441
 have properties different from those expected of
 metal-rich field RRLs.  In this paper we consider new
 observations of RRLs in both clusters which indicate that
 NGC~6388 and NGC
 6441 do not fit into either the Oo~I or Oo~II groups.
 
 NGC~6388 and NGC~6441 are also of interest because observations
 obtained with the Hubble Space Telescope (Rich et al. 1997) showed that the HBs of the clusters
 slope upward as one goes toward the blue in a $V$,~\bv color-magnitude diagram.
 Theoretical simulations (Sweigart \& Catelan 1998) demonstrated that this
 slope cannot be caused by differences in age, mass loss on the red giant
 branch, or differential reddening. The latter possibility was also ruled
 out by Layden et al. (1999).  Sweigart \& Catelan found that theoretical
 scenarios which did explain this slope also predicted a high luminosity for RRLs
 in NGC~6388 and NGC~6441 (see also Sweigart 1999). Testing this prediction is an
 additional goal of
 this study.

 \section{Oosterhoff Classification of NGC~6388 and NGC~6441}
 
  New CCD observations of NGC~6388 and NGC~6441 were obtained in 1998 with
  the 0.9-m telescope at Cerro Tololo Interamerican Observatory. These
  observations were used to approximately double the number of previously
  known RRLs within each cluster, and to determine periods and $B$,~$V$ light
  curves for new and previously known variables.  Details of the observations
  are presented in Pritzl et al. (1999a, 1999b)\markcite{P99a}\markcite{P99b}.
  Those papers also deal with the question of RRL membership in the two
  clusters.  Membership for RRab stars is usually straightforward but,
  as Layden et al. (1999) also found, there can be occasional confusion
  for RRc stars. For the purposes of this paper,
  we include only variables for which the case for membership is
  strong.  In particular, several possible shorter period RRc are not
  included in deriving the $N_{\rm c}/N_{\rm RR}$ ratios of Table~1.
  The more questionable variables are dealt with fully
  in Pritzl et al. (1999a, 1999b).

\vskip 0.15in
\parbox{3in}{\epsfxsize=3.25in \epsfysize=1.5in \epsfbox{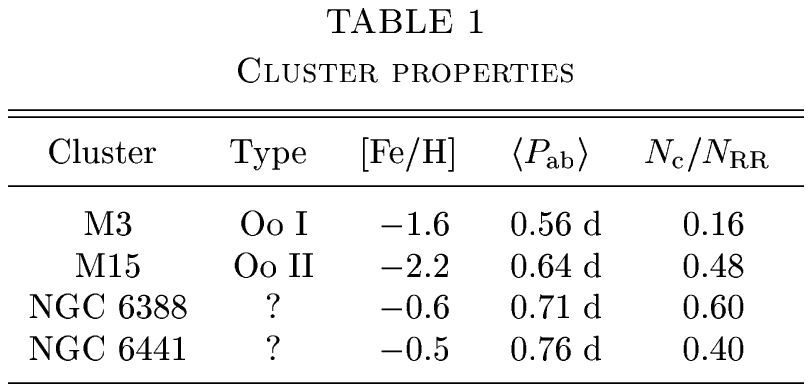}} 
\vskip 0.15in

  Mean properties of RRLs in NGC~6388 and NGC~6441 are summarized in
  Table~1, together with those of RRLs in M3 and M15, typical Oo~I and Oo~II
 clusters.
  NGC~6388 and NGC~6441 are distinguished by
  the surprisingly long mean
  periods of their RRab stars.  The distinction is emphasized in
  Figure~1, where NGC~6388 and NGC~6441 stand out sharply from Oo~I and
  Oo~II clusters in the $\langle P_{\rm ab} \rangle$
  versus [Fe/H] diagram.  NGC
 6388 and NGC~6441 are not only the most metal-rich clusters plotted in
 Figure~1, but
  also the clusters with the largest values of
  $\langle P_{\rm ab} \rangle$, completely contradicting the trend
  seen among the other clusters.

\vskip 0.2in
\parbox{3in}{\epsfxsize=3.25in \epsfysize=3.0in \epsfbox{FeH.eps}} 
\vskip 0.1in
\centerline{\parbox{3.5in}{\footnotesize  {\sc Fig.~1.---}Mean period vs. 
 [Fe/H] diagram showing the
 offset of NGC~6388 (circle)
 and NGC~6441 (square) from the Oosterhoff~I (crosses) and Oosterhoff~II
 (asterisks) globular clusters.  Data for the Oosterhoff clusters are taken
 from Sandage (1993b).
\label{fig1}         
       }}
\vskip 0.2in

  The unusual nature of NGC~6388 and NGC~6441 is also indicated in the
  period-amplitude diagram plotted as Figure~2. The usual differences between
  the Oosterhoff groups are apparent in this figure.
  At constant amplitude, RRab stars in the Oo~II clusters M15
 (Silbermann et al. 1995\markcite{S95}; Bingham et al. 1984\markcite{B84})
 and M68 (Walker 1994\markcite{W94}) are shifted
  toward longer periods
 compared to those in the Oo~I cluster M3 (Carretta et al. 1998\markcite{Ca98}),
 while metal-rich field RRab
 stars occur at shorter periods.  In contrast,
 NGC~6388 and NGC~6441
 RRab stars have, at a given amplitude, periods as long as, and in some
 cases, longer than those
 of Oo~II RRab stars. It should be noted that, in selecting comparison
 stars to
 plot in Figure~2, obvious Blazhko variables have been excluded, but no
 more stringent light curve criteria have been applied.

\vskip 0.25in
\parbox{3in}{\epsfxsize=3.25in \epsfysize=3.0in \epsfbox{PA.eps}} 
\vskip 0.1in
\centerline{\parbox{3.5in}{\footnotesize  {\sc Fig.~2.---}Period-amplitude 
 diagram for the ab-type RR Lyrae variables of NGC~6388 (open circles) 
 and NGC~6441 (filled circles) as compared to field
 RR Lyrae of ${\rm [Fe/H]} \ge -0.8$ (asterisks), V9 in 47~Tuc (six pointed
 star), M3 (open boxes), M15 (stars),
 and M68 (triangles).  The smaller circles denote variables that are
 believed to be blended with companions or possibly to be Blazhko stars. 
       }}
\vskip 0.25in

 RRab star periods longer than
 0.8~days account for 50 and 37 percent of the RRab stars in
 NGC~6388
 and NGC~6441 respectively.
 Such long periods are rare but not unprecedented among
 other globular
 clusters. The globular cluster $\omega$~Centauri, unique in containing
 RRLs with a wide range in [Fe/H] (Butler, Dickens, \& Epps 1978\markcite{B78}),
 also contains a
 significant number of very long period RRab stars.  Although
 $\omega$~Cen is primarily an Oo~II cluster, it has been
 suggested that it contains RRLs belonging to both Oosterhoff groups
 (Butler et al. 1978\markcite{B78}).  However, most of its RRab stars have
 periods much shorter than those in NGC~6388 and NGC~6441. As another example,
 Wehlau (1990\markcite{W90}) found the three RRab stars in the globular cluster
 NGC~5897 to all have periods
 longer than 0.79~d. NGC~5897 is a metal-poor cluster, however, with
 ${\rm [Fe/H]} = -1.68$ (Zinn \& West 1984\markcite{zw84}), and in that
 regard is unlike NGC~6388 and NGC~6441. With a period of 0.737~d, the
 RRL V9 in the metal-rich globular cluster 47~Tuc may be a closer analogue
 to the RRab stars in NGC~6388 and NGC~6441 (Figure~3 of Sweigart
 \& Catelan 1998; Carney et al. 1993\markcite{C93}).
 
 Period histograms for NGC~6388, NGC~6441, M3, and M15 are shown in
 Figure~3.  Like the very metal-poor Oo~II clusters,
 NGC~6388 and NGC~6441 are relatively rich in RRc stars.  On the other hand,
 as we have already noted, the metallicites of NGC~6388 and NGC~6441 are
 similar to, but even higher than, those of Oo~I.
 Once again, NGC~6388 and NGC~6441 stand out as anomalous.
 
 We conclude that NGC~6388 and NGC~6441 cannot be readily classified as either
 Oo~I or Oo~II from the properties of their RRLs.  The long mean periods of their
 RRab stars, their location in the period-amplitude diagram, and the large
 proportions
 of RRc stars all support an Oo~II classification (see also
 Clement 1999a).  However, the mean RRab
 periods
 are longer than for Oo~II clusters and the high metallicities of NGC~6388
 and NGC~6441 stand in contradiction to the low metallicities of Oo~II systems.
 We also note that NGC~6388 and NGC~6441 are very different from the globular
 clusters of the Large Magellanic Cloud, which do not fall into either Oo group
 (Bono, Caputo, \& Stellingwerf 1994).  Those clusters are metal-poor and have
 values of
 $\langle P_{\rm ab} \rangle$ intermediate between Oo~I and Oo~II.  We therefore
 suggest that NGC~6388 and NGC~6441 might represent a new Oosterhoff class.

 \section{The luminosity of the RR Lyrae Stars}
 
 Sandage et al. (1981)\markcite{SKS81} noted a shift in period
 between RRLs in M3 and M15, measured at constant $T_{\rm eff}$ or
 constant amplitude. Using Ritter's relation,
 $P\sqrt\langle\rho\rangle = Q$,
 they interpreted this as evidence that the M15 RRLs were less dense and thus
 more luminous than
 those in M3.  This was later generalized to a luminosity-metallicity
 correlation, in the sense that RRL brightness increases with decreasing
 [Fe/H] (Sandage 1982\markcite{S82}; Carney et al. 1992\markcite{C92}).
 This luminosity-metallicity
 correlation is generally represented by a linear equation of the form
 $M_V = \alpha \times {\rm [Fe/H]} + \beta$, where $\alpha$ denotes the
 sensitivity of RRL luminosity to metallicity. The size of
 this correlation remains subject to debate, with values of $\alpha$ ranging
 from 0.3 (Sandage 1993a) to 0.13 (Fusi Pecci et al. 1996). We note,
 however, that there has been recognition that this linear relationship
 may not always apply, as, for example, in describing the relationship
 between metallicity and luminosity among the RR Lyrae variables in
 $\omega$ Cen (Butler et al. 1978; Dickens 1989; Lee 1991).
 
 On the basis of this prior work, one would expect the locus of RRab stars
 in the period-amplitude diagram
 to shift to shorter periods with increasing [Fe/H].  Comparison of the locations
 in the period-amplitude diagram
 of RRab stars in the very metal-poor globular clusters M15 and M68 with
 RRab stars in M3 and with metal-rich field RRab stars (Figure~2) confirms
 this expectation.  On the other hand, RRab stars in
 NGC~6388 and NGC~6441 are shifted toward longer periods than would be expected
 from
 their metallicities, indicating that they are at least as bright as RRLs in
 very metal-poor Oo~II clusters.
 
 Sweigart \& Catelan (1998)\markcite{SC98}, in an effort to explain the
 unusual slope of the HBs of these two clusters,
 created three theoretical scenarios which could be tested using RRLs. Their
 models predict that the blue HBs of NGC~6388 and NGC~6441 should be
 unusually bright. Though at the time, the data available on the RRLs
 of the two clusters were slight, available observations were
 consistent with the predictions of these models.
 
 The data now available make the case much more strongly. The boxed area in
 Figure~2
 represents one of the model predictions (helium-mixing scenario)
 of Sweigart \& Catelan [1998, from
 their Figure~3 as translated to the period-amplitude diagram by Layden et al.
 (1999, cf. their Figure~9)]. The period-amplitude data are in similarly
 good agreement with the other scenarios of Sweigart \& Catelan, all of
 which require that the RRLs of NGC~6388 and NGC~6441 are brighter than
 solar neighborhood field RRLs of comparable [Fe/H].

\begin{figure*}[ht]
\figurenum{3} 
 \centerline{\epsfig{file=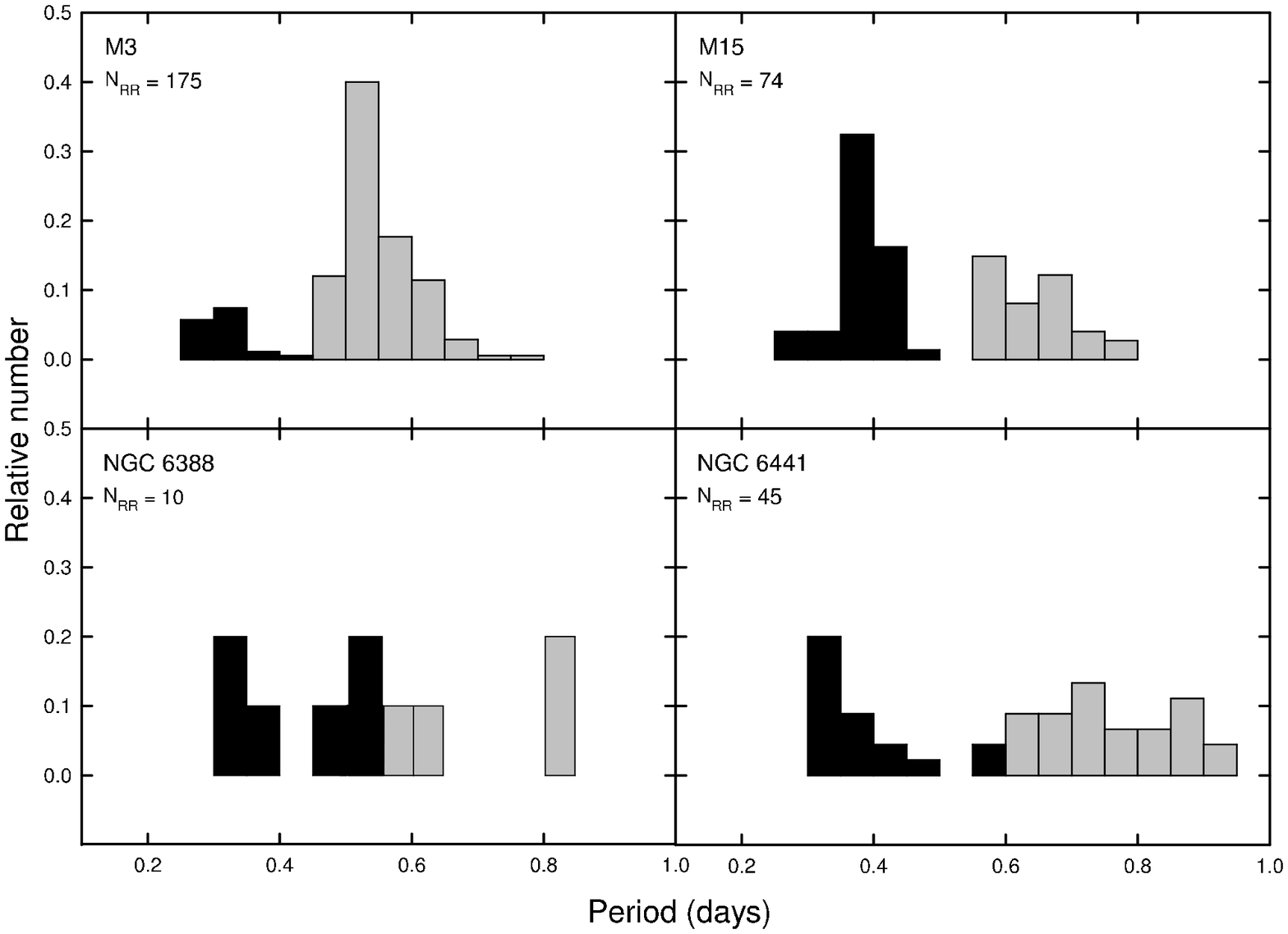,height=3.85in,width=5.25in}}
 \caption{Period distribution histograms for the RR Lyraes
  of M3, M15, NGC~6388, and
  NGC~6441.  The darker area is occupied by RRc variables. The lighter area 
  is occupied by RRab variables.  Data for M3 and M15 are taken from
  Clement (1999b).  
 \label{fig3}
         }
\end{figure*}

 It has been argued that HB evolution, rather than metallicity {\it per
 se}, might be the governing factor in determining whether a cluster belongs
 to Oo~I or Oo~II (Clement \& Shelton 1999\markcite{C99}; Lee \& Carney
 1999\markcite{LC99}).  Lee et al. (1990) also argued that evolution was
 an important element in the origin of the Oosterhoff phenomenon.
 Oo~II clusters usually have bluer HBs than Oo~I clusters, although
 there are exceptions such as M28 or NGC 4147 (cf. Table~1 in Castellani
 \& Quarta 1987\markcite{CQ87}). According to this explanation RRLs in Oo~II
 clusters spend most of their HB lifetimes on the blue HB (BHB) before evolving
 redward through the instability strip on their way back to the
 asymptotic-giant branch.  The final crossing of the instability strip occurs at a
 brighter luminosity and hence longer period than for stars near the ZAHB.
 
 NGC~6388 and NGC~6441 have
 predominantly red HBs with pronounced blue components. The sloping HB morphology
 in
 the color-magnitude diagrams of NGC~6388 and NGC~6441 does not indicate
 that the RRLs in those clusters have evolved from the BHB.
 Moreover, Sweigart \& Catelan's (1998) models indicate that the RRLs are in
 the main phase of HB evolution, requiring that the HBs of the clusters are
 unusually bright.  Thus, evolution does not appear to be the explanation for the
 long RRL periods in NGC~6388 and NGC~6441.
 
 The location of the NGC~6388 and NGC~6441 RRab stars in the period-amplitude
 diagram leads us to conclude that RRLs in those metal-rich globular clusters
 are at least as bright as those in the very metal-poor Oo~II clusters M15 and
 M68.  There is thus no universal correlation between RRL luminosity and
 metallicity.
 The HB morphology of the two clusters, together with the theoretical
 scenarios of Sweigart \& Catelan, further lead us to conclude that the bright
 RRLs are a consequence of bright HBs rather than evolution from the BHB.

  \section{Discussion}
 
  The relatively metal-rich globular clusters NGC~6388 and NGC~6441 are
  distinct in several ways from ordinary Oo~I and Oo~II clusters.
  Nor are their RRLs similar to those of the metal-rich
  field population of the solar neighborhood.  The location of NGC~6388
  and NGC~6441 RRab stars
  in the period-amplitude diagram
  is consistent with the RRLs of the two clusters being
  as bright or slightly brighter than those of Oo~II clusters such
  as M15 or M68, a result consistent with the theoretical models of
  Sweigart \& Catelan (1998). The RRLs in NGC~6388 and NGC~6441 thus
  demonstrate that RRL luminosity is not always inversely correlated with
  metallicity.
 
  Should we then regard NGC~6388 and
  NGC~6441 as sufficiently distinct from Oo~I and Oo~II
  clusters to be representatives of a third Oosterhoff group?  Or
  should we instead regard them as an aberrant type of Oo~II cluster?
  It is to some degree a matter of semantics,
  the answer depending in part upon which characteristics one regards
  as essential to Oosterhoff classification.  It is
  nonetheless worth noting that
  NGC~6388 and NGC~6441 are alike in ways other than [Fe/H] and the
  properties of their RRLs.  Both are among the most luminous globular
  clusters of the Galaxy and both have very high central densities.
  It remains an open but intriguing question whether those
  attributes play a role in
  producing the unusual RRL populations of the clusters.
 
 \begin{acknowledgements}
 
 This work has been supported by the National Science Foundation under
 grant AST95-28080.
 Support for M.C. was provided by NASA through Hubble Fellowship grant
 HF--01105.01--98A awarded by the Space Telescope Science Institute,
 which is operated by the Association of Universities for Research in
 Astronomy, Inc., for NASA under contract NAS~5-26555.
 \end{acknowledgements}

 \end{document}